\documentclass[12pt]{article} 

\usepackage{epsfig} 
\usepackage{amsbsy}  
 
\newlength{\largfig}
\def\ds#1{#1\kern-1ex\hbox{/}} 
\def\sl#1{#1\kern-1ex\hbox{/}} 
\def\dsh{h\kern-1.2ex /}

\def\beq{\begin{equation}} 
\def\eeq{\end{equation}} 
\def\eq{\beq\eeq} 
\def\beqn{\begin{eqnarray}} 
\def\eeqn{\end{eqnarray}} 
 
\def\lq{\left[} 
\def\rq{\right]} 
\def\rg{\right\}} 
\def\lg{\left\{} 
\def\({\left(} 
\def\){\right)} 
 
\def\ba{\begin{eqnarray}} 
\def\ea{\end{eqnarray}} 
\def\bq{\begin{equation}} 
\def\eq{\end{equation}}

\def\sla#1{\ifmmode% 
\setbox0=\hbox{$#1$}% 
\setbox1=\hbox to\wd0{\hss$/$\hss}\else% 
\setbox0=\hbox{#1}% 
\setbox1=\hbox to\wd0{\hss/\hss}\fi% 
#1\hskip-\wd0\box1 } 
 
\def\sing{{\rm sing}}
\def\qb{{\bar q}}
\def\MB{{\cal M}_B}

\def\I{{\cal I}} 
\def\I{{\boldsymbol{I}}}

\def\asb{{}\ifmmode \bar{\alpha}_s \else $\bar{\alpha}_s$\fi} 
\def \as   {\ifmmode \alpha_s \else $\alpha_s$ \fi}

\def\tag{\rm tag}

\def\so3#1{\,{\rm S}_{1,\,3}\left(#1 \right)} 
\def\st2#1{\,{\rm S}_{2,\,2}\left(#1 \right)} 

\def\Re{\mathop{\rm Re}}

\newskip\humongous \humongous=0pt plus 1000pt minus 1000pt

\newif\ifdtup

\catcode`@=12 
 
\catcode`\@=11 
 
\@addtoreset{equation}{section} 

\def\theequation{\thesection.\arabic{equation}} 
 
\def\@normalsize{\@setsize\normalsize{15pt}\xiipt\@xiipt 
\abovedisplayskip 14pt plus3pt minus3pt% 
\belowdisplayskip \abovedisplayskip 
\abovedisplayshortskip \z@ plus3pt% 
\belowdisplayshortskip 7pt plus3.5pt minus0pt} 
 
\def\small{\@setsize\small{13.6pt}\xipt\@xipt 
\abovedisplayskip 13pt plus3pt minus3pt% 
\belowdisplayskip \abovedisplayskip 
\abovedisplayshortskip \z@ plus3pt% 
\belowdisplayshortskip 7pt plus3.5pt minus0pt 
\def\@listi{\parsep 4.5pt plus 2pt minus 1pt 
     \itemsep \parsep 
     \topsep 9pt plus 3pt minus 3pt}} 
 
\@twosidetrue

\catcode`\@=11  
\def\section{\@startsection{section}{1}{\z@}{3.5ex plus 1ex minus 
   .2ex}{2.3ex plus .2ex}{\large\bf}}

\def\thesection{\arabic{section}} 
\def\thesubsection{\arabic{section}.\arabic{subsection}} 
\def\thesubsubsection{\arabic{section}.\arabic{subsection}.\arabic{subsubsection}} 
 
\def\appendix{\setcounter{section}{0} 
 \def\thesection{\Alph{section}} 
 \def\theequation{\Alph{section}.\arabic{equation}} 
 \def\thesubsection{\Alph{section}.\arabic{subsection}} 
\def\thesubsubsection{\Alph{section}.\arabic{subsection}.\arabic{subsubsection}} 
 
\def\section{\@startsection{section}{1}{\z@}{3.5ex plus 1ex minus 
   .2ex}{2.3ex plus .2ex}{\large\bf}} 
}

\newcommand{\ccaption}[2]{ 
  \begin{center} 
    \parbox{0.85\textwidth}{ 
      \caption[#1]{\small\it {#2}}} 
  \end{center}    } 
 
\def \ep{\epsilon} 
 
\def \to   {\mbox{$\rightarrow$}}

\newcount\minutes 
\newcount\scratch 
 
\def\timestamp{% 
\scratch=\time 
\divide\scratch by 60 
\edef\hours{\the\scratch} 
\multiply\scratch by 60 
\minutes=\time 
\advance\minutes by -\scratch 
---$\,$\hours:\null 
\ifnum\minutes< 10 0\fi 
\the\minutes}

\begin{document} 
\begin{titlepage} 
\nopagebreak 
{\flushright{ 
        \begin{minipage}{5cm} 
	 DCPT/03/54\\
	 IPPPP/03/27 \\
         MADPH 03-1330 \\ 	 
        {\tt hep-ph/0306109}\hfill \\ 
        \end{minipage}        } 
 
} 
\vfill 
\begin{center} 
{\LARGE \bf \sc 
 \baselineskip 0.9cm 
Next-to-leading order jet distributions for Higgs boson production via \\
weak-boson fusion 

} 
\vskip 0.5cm  
{\large   
T.~Figy$^1$, C.~Oleari$^2$ and D.~Zeppenfeld$^1$ 
}   
\vskip .2cm  
{$^1$ {\it Department of Physics, University of Wisconsin, Madison, WI 
53706, U.S.A. }}\\   
{$^2$ {\it Department of Physics, University of Durham,
South Road, Durham DH1 3LE, U.K.}}\\

\vskip 
1.3cm     
\end{center} 
 
\nopagebreak 
%\vfill 
%\vskip 3cm 
\begin{abstract}
The weak-boson fusion process is expected to provide crucial information on
Higgs boson couplings at the Large Hadron Collider at CERN. The achievable
statistical accuracy demands comparison with next-to-leading order QCD
calculations, which are presented here in the form of a fully flexible 
parton Monte Carlo program. 
QCD corrections are determined for jet distributions and are
shown to be modest, of order 5 to 10\% in most cases, but reaching 30\% 
occasionally. Remaining scale uncertainties range
from order 5\% or less for distributions to below $\pm 2$\% for the Higgs 
boson cross section in typical weak-boson fusion search regions.
\end{abstract} 
\vfill 
%\today \timestamp \hfill 
\vfill 
\end{titlepage} 
\newpage

\section{Introduction} 
\begin{figure}[t] 
\centerline{ 
\epsfig{figure=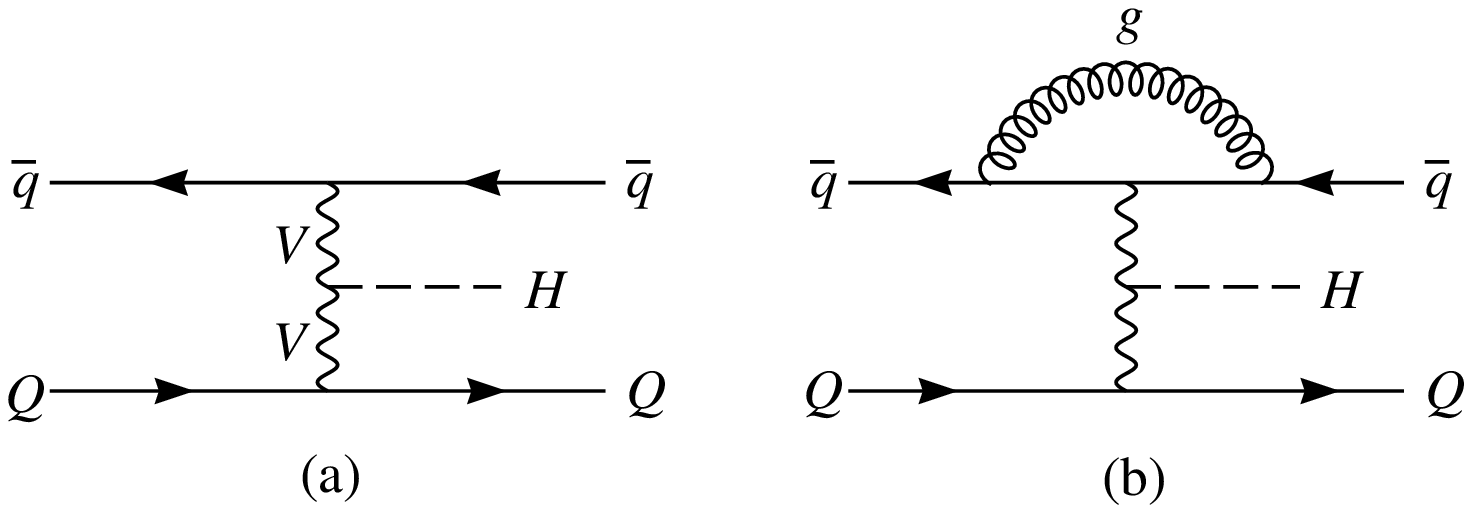,width=0.8\textwidth,clip=} \ \  
} 
\ccaption{} 
{ \label{fig:feyn2} Feynman graphs contributing to $\bar qQ\to \bar qQH$ at
(a) tree level and (b) including virtual corrections to the upper quark line.
}
\end{figure}

The weak-boson fusion (WBF) process, $qQ\to qQH$,
is expected to provide a copious
sources of Higgs bosons in $pp$-collisions at the Large Hadron 
Collider (LHC) at CERN. It can be visualized (see Fig.~\ref{fig:feyn2})
as the elastic scattering 
of two (anti-)quarks, mediated by $t$-channel $W$ or $Z$-exchange, with the 
Higgs boson radiated off the weak-boson propagator. Together with gluon fusion,
it represents the most promising production process for Higgs boson
discovery~\cite{CMS,ATLAS}. 
Once the Higgs boson has been found and its mass determined, the 
measurement of its couplings to gauge bosons and fermions will be of 
main interest~\cite{Zeppenfeld:2000td}.
Here WBF will be of central importance since it allows for independent 
observation in the $H\to\tau\tau$~\cite{wbfhtautau}, $H\to WW$~\cite{wbfhtoww},
$H\to\gamma\gamma$~\cite{wbfhtophoton} and 
$H\to$~invisible~\cite{Eboli:2000ze} channels. This multitude of 
channels is crucial for separating the effects of different Higgs boson
couplings.

The WBF measurements can be performed at the LHC with statistical accuracies
on cross sections times decay branching ratios, $\sigma\cdot B$, reaching
5 to 10\%~\cite{Zeppenfeld:2000td}. In order to extract Higgs boson
coupling constants with this full statistical power, a theoretical 
prediction of the Standard Model (SM) production cross section with error
well below 
10\% is required, and this clearly entails knowledge of the next-to-leading
order (NLO) QCD corrections. 

For the total Higgs boson production cross section via WBF these NLO
corrections  
have been available for a decade~\cite{WBF_NLO} and they are relatively
small, with $K$-factors around 1.05 to 1.1. These modest $K$-factors are
another reason for the importance of Higgs boson production via WBF: 
theoretical uncertainties will not limit the precision of the coupling 
measurements. This is in contrast to the dominant gluon fusion channel
where the $K$-factor is larger than 2 and residual uncertainties of 
10-20\% remain, even after the 2-loop corrections have been 
evaluated~\cite{HggNLO,H2loop}.

In order to distinguish the WBF Higgs boson signal from backgrounds, stringent 
cuts are required on the Higgs boson decay products as well as on the two
forward 
quark jets which are characteristic for WBF. Typical cuts have an acceptance
of less than 25\% of the starting value for $\sigma\cdot B$. The question then
arises whether the $K$-factors and the scale dependence determined for the
inclusive production cross section~\cite{WBF_NLO} are valid for the Higgs
boson search region also. This is best addressed by implementing the 
one-loop QCD
corrections in a fully flexible NLO parton-level Monte Carlo program.

We are presently developing such programs for a collection of relevant WBF 
processes, of which Higgs boson production, in the narrow resonance 
approximation, is the simplest example. The purpose of this paper then is 
twofold. First we use the Higgs boson signal process as our example to discuss
the generic features of NLO QCD corrections to WBF processes. We use the 
subtraction method of Catani and Seymour~\cite{CS} throughout. In 
Section~\ref{sec:subtraction} we describe the handling of real emission 
singularities. We give explicit formulas for the finite contributions 
which remain after factorization of the initial-state collinear singularities
and after cancellation of divergences produced by soft and collinear 
final-state gluons against the corresponding terms in the virtual corrections. 

This procedure yields a regularized Monte Carlo program which allows us to 
determine infrared safe observables at NLO. The main features of the program,
numerical tests, and parameters to be used in the later phenomenological 
discussion are described in Section~\ref{sec:NLOMC}.
In Section~\ref{sec:pheno} we then
use this tool to address our second objective, a discussion of the 
QCD radiative corrections as a function of jet observables. We determine
the $K$-factors and the residual scale uncertainties for distributions of 
the tagging jets which represent the scattered quarks in WBF. In addition,
we quantify the cross section error induced by uncertainties in the 
determination of parton distribution functions (pdf's). Pdf errors and 
scale variations in the phase-space regions relevant for the Higgs boson 
search turn out to be quite small (approximately 4\% when combined) and 
thus indicate the small theoretical uncertainties
required for reliable coupling measurements. Conclusion are presented in 
Section~\ref{sec:conclusions}.

\section{Subtraction terms for soft and collinear radiation}
\label{sec:subtraction}

At lowest order, Higgs boson production via weak-boson fusion is represented
by a single Feynman graph, like the one depicted in Fig.~\ref{fig:feyn2}(a)
for $\bar qQ\to \bar qQH$. We use this particular process to describe the
QCD radiative corrections. Generalization to crossed processes ($\qb\to q$
and/or $Q\to \bar Q$) is straightforward. Strictly speaking, 
the single Feynman graph picture is valid for different quark flavors 
on the two fermion lines only. For identical flavors annihilation 
processes, like $\bar q q\to Z^*\to ZH$ with subsequent decay $Z\to \bar q q$
or similar $WH$ production channels,
contribute  as well. For $qq\to qqH$ or $\bar q\bar q\to \bar q\bar q H$
the interchange of identical quarks in the initial or final state needs to 
be considered in principal. However, in the phase-space regions where WBF
can be observed experimentally, with widely separated quark jets of very large
invariant mass, the interference of these additional graphs is strongly 
suppressed by large momentum transfer in the weak-boson propagators. Color
suppression further makes these effects negligible. In the following we 
systematically neglect any identical fermion effects.

\begin{figure}[t] 
\centerline{ 
\epsfig{figure=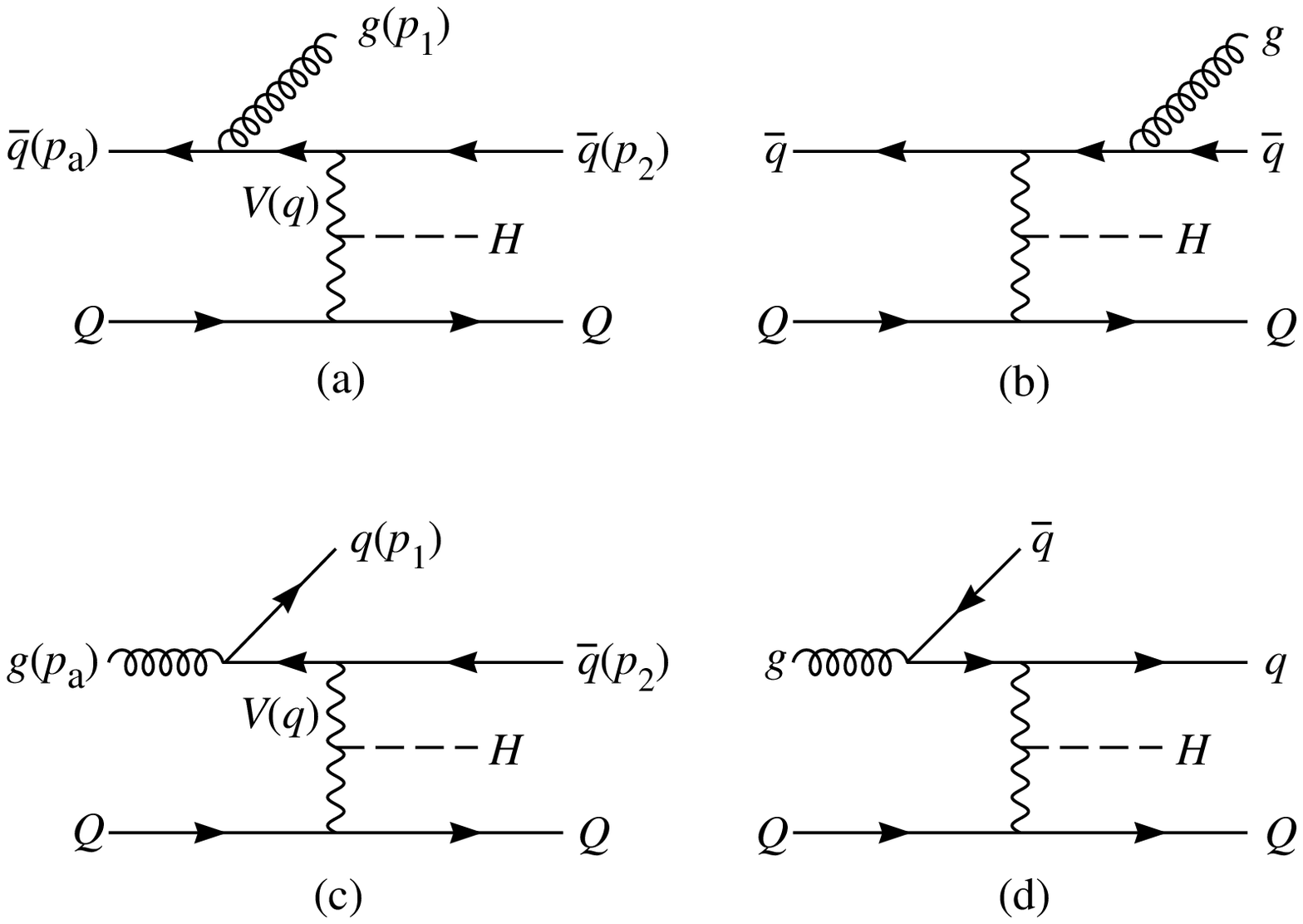,width=0.8\textwidth,clip=} \ \  
} 
\ccaption{} 
{ \label{fig:feyn3} Real emission contributions to Higgs boson production 
via weak-boson fusion. Corrections for the upper quark line only are shown:
gluon radiation ((a) and (b)) and gluon initiated processes ((c) and (d)).
}
\end{figure}

At NLO, the vertex corrections of Fig.~\ref{fig:feyn2}(b) and the real 
emission diagrams of Fig.~\ref{fig:feyn3} must be included. Because of 
the color singlet nature of the exchanged weak boson, any interference terms
between sub-amplitudes with gluons attached to both the upper and the lower
quark lines vanish identically at order $\alpha_s$. Hence it is sufficient to
consider radiative corrections to a single quark line only, which we here 
take as the upper one. Corrections to the lower fermion line are an exact copy.
We denote the amplitude for the real emission process
\beq
\bar q(p_a) + Q(p_b) \to g(p_1) + \bar q(p_2) + Q(p_3) + H(P)
\eeq
depicted in Fig.~\ref{fig:feyn3}(a) and (b) as 
${\cal M}^{\bar q}_r={\cal M}^{\bar q}_r(p_a,p_1,p_2;q)$, 
where $q=p_1+p_2-p_a$ is the 
four momentum of the virtual weak boson, $V$, of virtuality $Q^2=-q^2$.

The 3-parton phase-space integral of $|{\cal M}^{\bar q}_r|^2$ 
suffers from soft and collinear  divergences. They are absorbed in a single
counter term, which, in  
the notation of Ref.~\cite{CS}, contains the two dipole factors 
${\cal D}_2^{{\bar q} 1}$ and ${\cal D}_{12}^{\bar q}$
\beq        % checked 5/14
\label{eq:Mqsing}
\left|{\cal M}^{\bar q}\right|^2_\sing =  {\cal D}_2^{{\bar q} 1}+{\cal
  D}_{12}^{\bar q} 
 = 8\pi\alpha_s(\mu_R)\, C_F\, \frac{1}{Q^2}\,
     \frac{x^2+z^2}{(1-x)(1-z)} \left|\MB^\qb\right|^2\,,
\eeq
where $C_F=\frac{4}{3}$ and 
$\MB^\qb=\MB^\qb(\tilde p_a,\tilde p_2;q)$ is the Born amplitude 
for the lowest order process 
\beq
\bar q(\tilde p_a) + Q(p_b) \to \bar q(\tilde p_2) + Q(p_3) + H(P)\;,
\eeq
evaluated at the phase-space point 
\beq        % checked 5/14
\tilde p_a = x p_a\;,\qquad\qquad \tilde p_2 = p_1+p_2-(1-x)p_a\;,
\eeq
with
\beqn        % checked 5/14
\label{eq:defxz1}
      x &=& 1- \frac{p_1\cdot p_2}{(p_1+p_2)\cdot p_a} \,, \\
      z &=& 1-\frac{p_1\cdot p_a}{(p_1+p_2) \cdot p_a} =
              \frac{p_2\cdot p_a}{(p_1+p_2) \cdot p_a}\,.
\label{eq:defxz}
\eeqn
This choice continuously interpolates between the singularities due to
final-state soft gluons ($p_1\to 0$ corresponding to $x\to 1$ and $z\to 1$), 
collinear final-state partons ($p_1||p_2$ resulting in $p_1\cdot p_2\to 0$ 
or $x\to 1$) and gluon emission collinear to the initial-state anti-quark 
($p_1\to (1-x)p_a$ and $z\to 1$). The subtracted real emission 
amplitude squared, $|{\cal M}^\qb_r|^2-|{\cal M}^\qb|^2_\sing$, leads to a 
finite phase-space integral of the real parton emission cross section
\beqn        % checked 5/14
\sigma_3^{NLO}\(\qb Q\to \qb QHg\) \!\!\! &=& \!\!\!\int_0^1dx_a\int_0^1dx_b 
\, f_{\qb/p}\(x_a,\mu_F\) \, f_{Q/p}\(x_b,\mu_F\)\, 
\frac{1}{2\hat s} \, d\Phi_4\(p_1,p_2,p_3,P;p_a+p_b\)\nonumber \\
\label{eq:sig3nlo} 
&&
\times \lg \left|{\cal M}^\qb_r\right|^2F_J^{(3)}\(p_1,p_2,p_3\)-
      \left|{\cal M}^\qb\right|^2_\sing F_J^{(2)}\(\tilde p_2,p_3\) \rg \;,
\eeqn
where $\hat s = (p_a+p_b)^2$ is the center-of-mass energy.
The functions $F_J^{(3)}$ and $F_J^{(2)}$ define the jet algorithm for
3-parton and 2-parton final states and we obviously need 
$F_J^{(3)}\to F_J^{(2)}$ in the singular limits discussed above, i.e.\ the
jet algorithm (and all observables) must be infrared and collinear safe.
Being finite, the phase-space integral of Eq.~(\ref{eq:sig3nlo})
is evaluated numerically in $D=4$ dimensions.
Similarly, for the gluon initiated process 
\beq
g(p_a) + Q(p_b) \,\to\, q(p_1) + \bar q(p_2)+ Q(p_3) + H(P)\;,
\eeq
the singular behavior for $g\to q\qb$ splitting is absorbed into the 
singular counter term
\beqn        % checked 5/14
      \left|{\cal M}^g\right|^2_{\sing} = {\cal D}^{g1}_2+{\cal D}^{g2}_1
%\nonumber \\
      &=& 8\pi\alpha_s(\mu_R) \,T_F \,\frac{1}{Q^2}\,
      \Biggl[\frac{x^2+(1-x)^2}{1-z} \left|\MB^\qb\(\tilde p_a,\tilde
      p_2;q\)\right|^2  
\nonumber \\ \label{eq:ctg}
      &&\phantom{8\pi\alpha_s(\mu_R) \,T_F \,\frac{1}{Q^2}\,\Biggl[} +
\frac{x^2+(1-x)^2}{z} \left|\MB^q\(\tilde p_a,\tilde p_2;q\)\right|^2\Biggr]
  \; ,
\eeqn
where $T_F=\frac{1}{2}$ and 
$\MB^\qb$ and $\MB^q$ denote the Born amplitudes for the leading-order (LO)
processes 
$\bar q(\tilde p_a) + Q(p_b) \to \bar q(\tilde p_2) + Q(p_3) + H(P)$ and 
$q(\tilde p_a) + Q(p_b) \to q(\tilde p_2) + Q(p_3) + H(P)$, respectively.
The subtraction of $|{\cal M}^g|^2_{\sing}$ from the real emission amplitude
squared leads to a contribution to the subtracted 3-parton cross section
analogous to the one given in Eq.~(\ref{eq:sig3nlo}).

The singular counter terms are integrated analytically, in $D=4-2\epsilon$
dimensions, over the phase space of the collinear and/or soft final-state 
parton. Integrating Eq.~(\ref{eq:Mqsing}) yields the contribution 
(we are using the notation of Ref.~\cite{CS})  
\beq
\label{eq:I}
<\I(\ep)> = |\MB^{\bar q}|^2 \frac{\alpha_s(\mu_R)}{2\pi} C_F
\(\frac{4\pi\mu_R^2}{Q^2}\)^\epsilon \Gamma(1+\epsilon)
\lq\frac{2}{\epsilon^2}+\frac{3}{\epsilon}+9-\frac{4}{3}\pi^2\rq\;.
\eeq
We have regularized the divergences using dimensional reduction. If we
had used conventional dimensional regularization we would have obtained a
finite piece equal to $(10 - 4\pi^2/3)$.
The  $1/\epsilon^2$ and $1/\epsilon$
divergences cancel against the poles of the virtual 
correction, depicted in Fig.~\ref{fig:feyn2}(b). For the case at hand, the 
virtual correction amplitude ${\cal M}_V$ is particularly simple, leading 
to the divergent interference term
\beq         % checked 5/14
\label{eq:virtual_born}
2 \Re \lq {\cal M}_V\MB^* \rq
= |\MB^{\bar q}|^2 \frac{\alpha_s(\mu_R)}{2\pi} C_F
\(\frac{4\pi\mu_R^2}{Q^2}\)^\epsilon \Gamma(1+\epsilon)
\lq-\frac{2}{\epsilon^2}-\frac{3}{\epsilon}+c_{\rm virt}\rq\;.
\eeq
Here we have included the finite contribution of the virtual diagram which is
proportional to the Born amplitude. 
In dimensional reduction this contribution is given
by $c_{\rm virt}=\pi^2/3-7$ (
$c_{\rm virt}=\pi^2/3-8$
in conventional dimensional regularization). 

Summing together the contributions from Eq.~(\ref{eq:I}) and
Eq.~(\ref{eq:virtual_born}), we obtain the following finite 2-parton
contribution to the NLO cross section
\beqn        % checked 5/14
\sigma_2^{NLO}(\qb Q\to \qb QH) \!\!\!&=& \!\!\!\!\int_0^1\!dx_a\int_0^1\!dx_b
\, 
f_{\qb/p}\(x_a,\mu_F\) \, f_{Q/p}\(x_b,\mu_F\)\,
\frac{1}{2\hat s} \, d\Phi_3\(p_2,p_3,P;p_a+p_b\)\nonumber \\ \label{eq:sig2nlo}
& \times&\!\!\!\!\!
\left|\MB^\qb\right|^2\!F_J^{(2)}\!\(p_2,p_3\)\!
\lq 1+\frac{\alpha_s(\mu_{Ra})+\alpha_s(\mu_{Rb})}{2\pi}
\, C_F \( 9 - \frac{4}{3}\pi^2 +  c_{\rm virt}\)\rq\!.\phantom{aaaa}
\eeqn
The two $\alpha_s$ terms, at distinct renormalization scales $\mu_{Ra}$
and $\mu_{Rb}$, correspond to virtual corrections to the upper and the lower
fermion line in Fig.~\ref{fig:feyn2}, respectively, and we have anticipated 
the possibility of using different scales (like the virtuality of the attached
weak boson $V$) for the QCD corrections to the two fermion lines. 

The remaining divergent piece of the integral of the counter terms in
Eqs.~(\ref{eq:Mqsing}) and~(\ref{eq:ctg}) 
is proportional to the $P^{qq}$ and $P^{gq}$ splitting
functions and disappears after renormalization of the parton distribution
functions.  The surviving finite collinear terms are given by
\beqn        % checked 5/14
\sigma_{2,\rm coll}^{NLO}(\qb Q\to \qb QH) &=&
\int_0^1dx_a\int_0^1dx_b \,
f_{\qb/p}^c\(x_a,\mu_F,\mu_{Ra}\) \, f_{Q/p}\(x_b,\mu_F\)\; 
\nonumber \\ 
\label{eq:sig2coll}
&& \times \frac{1}{2\hat s} \, d\Phi_3\(p_2,p_3,P;p_a+p_b\)\,
\left|\MB^\qb\right|^2F_J^{(2)}\(p_2,p_3\) \;,
\eeqn
and similarly for quark initiated processes. Here the anti-quark 
function $f_{\qb/p}^c(x,\mu_F,\mu_R)$ is given by
\beqn        % checked 5/14
f_{\qb/p}^c(x,\mu_F,\mu_R)&=& \frac{\alpha_s(\mu_R)}{2\pi} 
\int_x^1 \frac{dz}{z}
\lg f_{g/p} \(\frac{x}{z},\mu_F\) A(z)\right.
\nonumber \\ 
&& + \left.
\lq f_{\qb/p} \(\frac{x}{z},\mu_F\)-z f_{\qb/p} \(x,\mu_F\) \rq B(z)
+f_{\qb/p}\(\frac{x}{z},\mu_F\) C(z) \rg
\nonumber \\ \label{eq:crossfct}
&& + \frac{\alpha_s(\mu_R)}{2\pi} f_{\qb/p} (x,\mu_F) D(x)\;,
\eeqn
with the integration kernels
\beqn        % checked 5/14
A(z) &=& T_F\lq z^2+(1-z)^2\rq\ln\frac{Q^2(1-z)}{\mu_F^2 z}
     +2T_F\; z(1-z)\;, \\
B(z) &=& C_F\lq
\frac{2}{1-z} \ln\frac{Q^2(1-z)}{\mu_F^2} -\frac{3}{2}\frac{1}{1-z}
\rq\;, \\
C(z) &=& C_F\lq 1-z-\frac{2}{1-z}\ln z - 
(1+z) \ln\frac{Q^2(1-z)}{\mu_F^2 z} \rq\;, \\
D(x) &=& C_F \lq \frac{3}{2} \ln\frac{Q^2}{\mu_F^2 (1-x)}
+2\ln(1-x)\ln\frac{Q^2}{\mu_F^2} +\ln^2(1-x) 
%- \frac{\pi^2}{3}-\frac{9}{2} -c_{\rm virt} \rq . \phantom{aa}
+ \pi^2 -\frac{27}{2} -c_{\rm virt} \rq . \phantom{aa}
\label{eq:dofx}
\eeqn
Note that $c_{\rm virt}$ exactly cancels between the contributions
from Eq.~(\ref{eq:sig2nlo}) and Eq.~(\ref{eq:dofx}). This fact will be used
below to numerically test our program.

The same kernels define the quark functions $f_{q/p}^c(x,\mu_F,\mu_R)$,
which appear with the Born amplitude $\MB^q(p_a,p_2;q)$ in the analog of
Eq.~(\ref{eq:sig2coll}) for the $qQ\to qQH$ processes. The gluon distribution
$f_{g/p}(x,\mu_f)$ thus appears twice, multiplying the Born amplitudes
squared $|\MB^q|^2$ and $|\MB^\qb|^2$ in the quark and anti-quark 
functions. These two terms correspond to the two terms in Eq.~(\ref{eq:ctg}),
after the $1/\epsilon$ collinear divergences have been factorized into 
the NLO parton distributions. 

Formulae identical to the ones given above for corrections to the upper 
line in the diagrams of
Fig.~\ref{fig:feyn3} apply to the case where the gluon is attached to the
lower line (with $a \leftrightarrow b$, $p_2 \leftrightarrow p_3$).  As for
the renormalization scale $\mu_R$ in Eq.~(\ref{eq:sig2nlo}), we 
distinguish  between the two factorization scales that appear for the upper
and lower quark lines, calling them $\mu_{Fa}$ and $\mu_{Fb}$, when needed.

A second class of gluon initiated processes arises
from crossing the final-state gluon and the initial-state quark $Q$ in the 
Feynman graphs of Fig.~\ref{fig:feyn3}(a) and (b). 
The resulting process can be described
as $g\qb\to \qb VH$ with the virtual weak boson $V$ undergoing the hadronic 
decay $V\to Q\bar Q$. Such contributions are part of the radiative corrections
to $\qb q\to VH$, they are suppressed in the WBF search regions with their
large dijet invariant mass, and we do not include them in our calculation.

%\newpage
\section{The NLO parton Monte Carlo program}
\label{sec:NLOMC}

The cross section contributions discussed above for the $\qb Q\to \qb QH$ 
process and crossing related channels have been implemented in a 
parton-level Monte Carlo program. The tree-level amplitudes are calculated 
numerically, using the helicity-amplitude formalism of Ref.~\cite{HZ}.
The Monte Carlo integration is performed with a modified version of 
VEGAS~\cite{vegas}.

The subtraction method requires the evaluation of real-emission amplitudes
and, simultaneously, Born amplitudes at related phase-space points (see
e.g.\ Eqs.~(\ref{eq:sig3nlo}) and (\ref{eq:sig2coll})). In order to speed 
up the program, the contributions from $\sigma_3^{NLO}$ and 
$\sigma_{2,\rm coll}^{NLO}$ are calculated in parallel, as part of the 
3-parton phase-space integral. Since the phase-space element
factorizes~\cite{CS}, 
\beq
\int d\Phi_4\(p_1,p_2,p_3,P;p_a+p_b\) = \int_0^1 dx \int_0^1 dz\,
d\Phi_3\(\tilde p_2,p_3,P;xp_a+p_b\) \frac{Q^2}{16\pi^2x} \;,
\eq
we can rewrite the finite collinear term of Eq.~(\ref{eq:sig2coll}) as
\beqn
\sigma_{2,\rm coll}^{NLO}\(\qb Q\to \qb QH\) &=& \int_0^1 dx_a\int_0^1 dx_b
\,\frac{1}{2(p_a+p_b)^2} \, d\Phi_4\(p_1,p_2,p_3,P;p_a+p_b\)
\phantom{aaaaaaaaaaa}
\nonumber \\ 
&& \times \Biggl\{ f_{g/p} (x_a,\mu_F) A(x) +
f_{\qb/p}(x_a,\mu_F)\Big[B(x)+C(x)\Big]
\nonumber \\ 
&& \phantom{\times \Biggl\{ }
+ x f_{\qb/p}(xx_a,\mu_F) \lq\frac{D(xx_a)}{1-xx_a} -B(x)\rq
\Biggr\}
\nonumber \\ \label{eq:sig2coll3}
&& \times f_{Q/p}\(x_b,\mu_F\)\; \frac{8\pi\alpha_s(\mu_R)}{Q^2}
    \left|\MB^\qb\right|^2 F_J^{(2)}\(\tilde p_2,p_3\) \;,
\eeqn
where $x$ and $z$ are determined as in Eqs.~(\ref{eq:defxz1})
and~(\ref{eq:defxz}).
Equation~(\ref{eq:sig2coll3}) allows for stringent consistency checks of our 
program, since we can determine the finite collinear cross section 
either as part of the 2-parton or as part of the 3-parton phase-space 
integral. For example, because of the cancellation of $c_{\rm virt}$ 
mentioned below Eq.~(\ref{eq:dofx}), the final result cannot depend on its
value. We have checked this independence numerically, at the $3\times 10^{-4}$
level. 
Another method to test the program is to determine the (anti-)quark 
functions $f^c_{a/p}(x,\mu_F,\mu_R)$ by numerical integration of 
Eq.~(\ref{eq:crossfct}), to then compute the finite collinear cross 
section together with the Born cross section, and to compare with the 
results of Eq.~(\ref{eq:sig2coll3}). For all phase-space regions considered,
our numerical program passes this test, with relative deviations of less
than $2\times 10^{-4}$ of the total Higgs boson cross section, which is the 
level of the Monte Carlo error.

As a final check we have compared our total Higgs boson cross section with 
previous analytical results~\cite{WBF_NLO}, as calculated with the 
program of Spira~\cite{Spira:1997dg}. We find agreement at or below the 
$1 \times 10^{-3}$ level 
which is inside the Monte Carlo accuracy for this comparison.

The cross sections to be presented below are based on CTEQ6M parton 
distributions~\cite{cteq6} with $\alpha_s(M_Z)=0.118$ for all NLO results 
and CTEQ6L1 parton distributions with $\alpha_s(M_Z)=0.130$ for all leading
order
cross sections. For all $Z$-exchange contributions the $b$-quark is included
as an initial and/or final-state massless parton. The $b$-quark contributions
are quite small, however, affecting the Higgs boson production cross section
at the 
$3$\% level only.
We choose $m_Z=91.188$~GeV, $\alpha_{QED}=1/128.93$ and the
measured value of $G_F$ as our electroweak input parameters from which we 
obtain $m_W=79.96$~GeV and $\sin^2\theta_W=0.2310$, using LO electroweak 
relations. 
In order to reconstruct jets from the final-state partons,
the $k_T$-algorithm~\cite{kToriginal} as described in Ref.~\cite{kTrunII} 
is used, with resolution parameter $D=0.8$.

\section{Tagging jet properties at NLO}
\label{sec:pheno}

The defining feature of weak-boson fusion events at hadron colliders
is the presence of two forward tagging jets, which, at LO, correspond to the
two scattered quarks in the process $qQ\to qQH$. Their observation, in 
addition to exploiting the properties of the Higgs boson decay products,
is crucial for the suppression of 
backgrounds~\cite{wbfhtautau,wbfhtoww,wbfhtophoton,Eboli:2000ze}. The 
stringent acceptance requirements imply that  tagging jet 
distributions must be known precisely for a reliable prediction of the 
SM Higgs signal rate. Comparison of the observed Higgs production rate
with this SM cross section, within cuts, then allows us to determine
Higgs boson couplings~\cite{Zeppenfeld:2000td} and, thus, the theoretical
error of the SM cross section directly feeds into the uncertainty of measured
couplings. 

The NLO corrections to the Higgs boson cross section do not depend on the 
phase space of the Higgs boson decay products because the Higgs boson, as a
scalar,                 
does not induce any spin correlations. It is therefore sufficient
to analyze tagging jet distributions to gain a reliable impression of the
size and the uncertainties of higher order QCD corrections.
Since search strategies depend on the decay mode considered and 
will evolve with time, we here consider generic weak-boson fusion cuts only.
They are chosen, however, to give a good approximation of the cuts 
suggested for specific Higgs boson search channels at the LHC. The phase 
space dependence of the QCD corrections and uncertainties, within these
cuts, should then provide a reasonably complete and reliable picture.

Using the $k_T$-algorithm, we calculate the partonic cross sections for events
with at least two hard jets,  which are required to have
\beq
\label{eq:cuts1}
p_{Tj} \geq 20~{\rm GeV} \, , \qquad\qquad |y_j| \leq 4.5 \, .
\eeq
Here $y_j$ denotes the rapidity of the (massive) jet momentum which is 
reconstructed as the four-vector sum of massless partons of 
pseudorapidity $|\eta|<5$. The Higgs boson decay products (generically 
called ``leptons'' in the following) are required to fall between the
two tagging jets in rapidity and they should be well observable. While 
an exact definition of criteria for the Higgs boson decay products will depend 
on the channel considered, we here substitute such specific requirements
by generating isotropic Higgs boson decay into two massless ``leptons'' 
(which represent $\tau^+\tau^-$ or $\gamma\gamma$ or $b\bar b$ final states)
and require
\beq
\label{eq:cuts2}
p_{T\ell} \geq 20~{\rm GeV} \,,\qquad |\eta_{\ell}| \leq 2.5  \,,\qquad 
\triangle R_{j\ell} \geq 0.6 \, ,
\eeq
where $R_{j\ell}$ denotes the jet-lepton separation in the rapidity-azimuthal
angle plane. In addition the two ``leptons'' are required to fall between the
two tagging jets in rapidity
\beq
\label{eq:cuts3}
y_{j,min}  < \eta_{\ell_{1,2}} < y_{j,max} \, .
\eeq
We do not specifically require the two tagging jets to reside in opposite 
detector hemispheres for the present analysis. Note that no reduction
due to branching ratios for specific final states is included in our 
calculation: the cross section without any cuts corresponds to the
total Higgs boson production cross section by weak-boson fusion.

At LO, the signal process has exactly two massless final-state quarks, which
are identified as the tagging jets, provided they pass the $k_T$-algorithm 
and the cuts described above. At NLO these jets may be composed of
two partons (recombination effect) or we may encounter three 
well-separated partons, which satisfy the cuts of Eq.~(\ref{eq:cuts1})
and would give rise to three-jet events. As with
LHC data, a choice needs to be made for selecting the tagging jets in such a
multijet situation. We consider here the following two possibilities:
\begin{enumerate}
\item[1)]
Define the tagging jets as the two highest $p_T$ jets in the event. This 
ensures that the tagging jets are part of the hard scattering event. We
call this selection the ``$p_T$-method'' for choosing tagging jets.

\item[2)]
Define the tagging jets as the two highest energy jets in the event. This 
selection favors the very energetic forward jets which are typical for 
weak-boson fusion processes. We call this selection the ``$E$-method'' for 
choosing tagging jets.
\end{enumerate}
Backgrounds to weak-boson fusion are significantly suppressed by requiring
a large rapidity separation of the two tagging jets. As a final cut, 
we require
\beq
\label{eq:cuts4}
\Delta y_{jj}=|y_{j_1}-y_{j_2}|>4\; ,
\eeq
which will be called the ``rapidity gap cut'' in the following.

\begin{figure}[t] 
\centerline{ 
\epsfig{figure=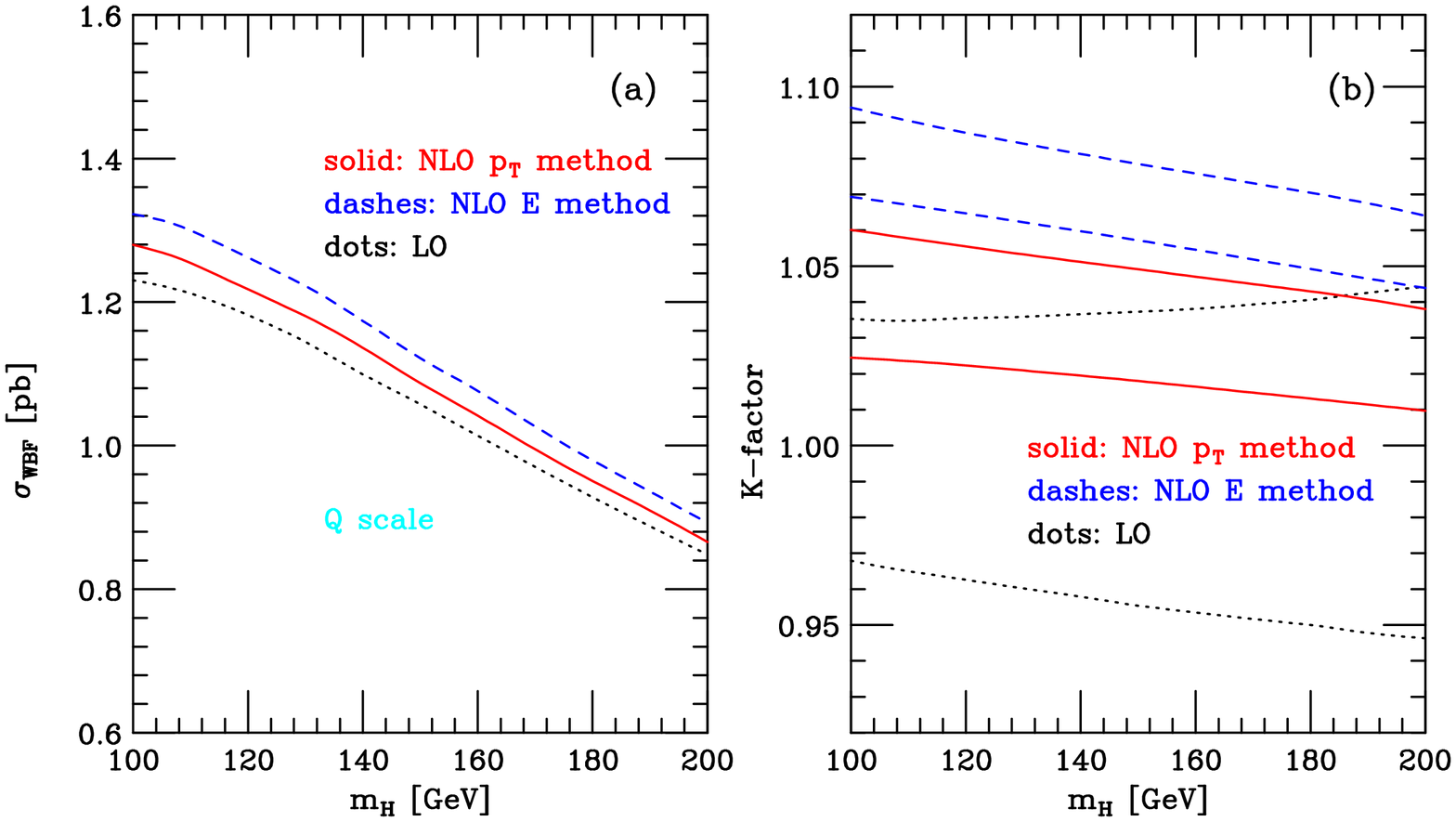,width=0.6\textwidth,angle=90,clip=} \ \  
} 
\ccaption{} 
{ \label{fig:sigtot} Effect of QCD radiative corrections on the Higgs boson
production cross section via WBF, as a function of the Higgs boson mass,
$m_H$. Results are given at LO (black dotted) and at NLO for
the $p_T$-method (solid red) and the $E$-method (dashed blue) for
defining tagging jets. Panel (a) gives the total cross section within the
cuts of Eqs.~(\protect\ref{eq:cuts1})--(\protect\ref{eq:cuts4}). 
The corresponding scale 
dependence, for variation of $\mu_R$ and $\mu_F$ by a factor of 2, is shown 
in panel (b). See text for details.
}
\end{figure} 

Cross sections, within the cuts of Eqs.~(\ref{eq:cuts1})--(\ref{eq:cuts4}), 
are shown in Fig.~\ref{fig:sigtot}(a), as a function of the Higgs boson 
mass, $m_H$. As for the total WBF cross section, the NLO effects are modest
for the cross section within cuts, amounting to a 3-5\% increase for the
$p_T$-method of selecting tagging jets (solid red) and a 6-9\% increase
when the $E$-method is used\footnote{The larger cross section for the 
$E$-method is due to events with a fairly energetic extra central
jet. A veto on central jets of $p_{Tj}>20$~GeV and rapidity between the two
tagging jets, as suggested for the WBF selection, lowers the NLO cross
section to $0.97\times\sigma_{LO}$ for the $p_T$-method and 
$0.93\times\sigma_{LO}$ for the $E$-method.}.
 These $K$-factors, and their scale dependence, 
are shown in Fig.~\ref{fig:sigtot}(b). Here the $K$-factor is defined as
\beq
\label{eq:Kfactor}
K=\frac{\sigma(\mu_R,\mu_F)}{\sigma^{LO}(\mu_F=Q_i)}\; ,
\eeq
i.e.\ the cross section is normalized to the LO cross section, determined
with CTEQ6L1 parton distributions, and a factorization scale which is set
to the virtuality of the weak boson which is attached to a given quark
line.

We have investigated two general scale choices. First we consider the
Higgs boson mass as the relevant hard scale, i.e.\ we set  
\beq
\label{eq:scale.mh}
\mu_F=\xi_F m_H\; ,\qquad \qquad \mu_R=\xi_R m_H\;.
\eeq
As a second option we consider
the virtuality of the exchanged weak boson. Specifically, independent 
scales $Q_i$ are determined for radiative correction on the upper and the 
lower quark line, and we set 
\beq
\label{eq:scale.Q}
\mu_{Fi}=\xi_F Q_i\; ,\qquad\qquad  \mu_{Ri}=\xi_R Q_i\;.
\eeq
This choice is motivated by the picture of WBF as two
independent DIS events, with independent radiative corrections on the
two electroweak boson vertices. In general we find the largest scale 
variations when we vary the renormalization scale and the factorization 
scale in the same direction. We only show results for this case, 
$\xi=\xi_R=\xi_F$, in the following.
The curves in Fig.~\ref{fig:sigtot}(b) correspond to the largest variations
found for $\xi=1/2$ and $\xi=2$ when considering both scale choices 
simultaneously. The residual scale uncertainty is about $\pm 5$\% at LO
and reduces to below $\pm 2$\% at NLO. 

\begin{figure}[t] 
\centerline{ 
\epsfig{figure=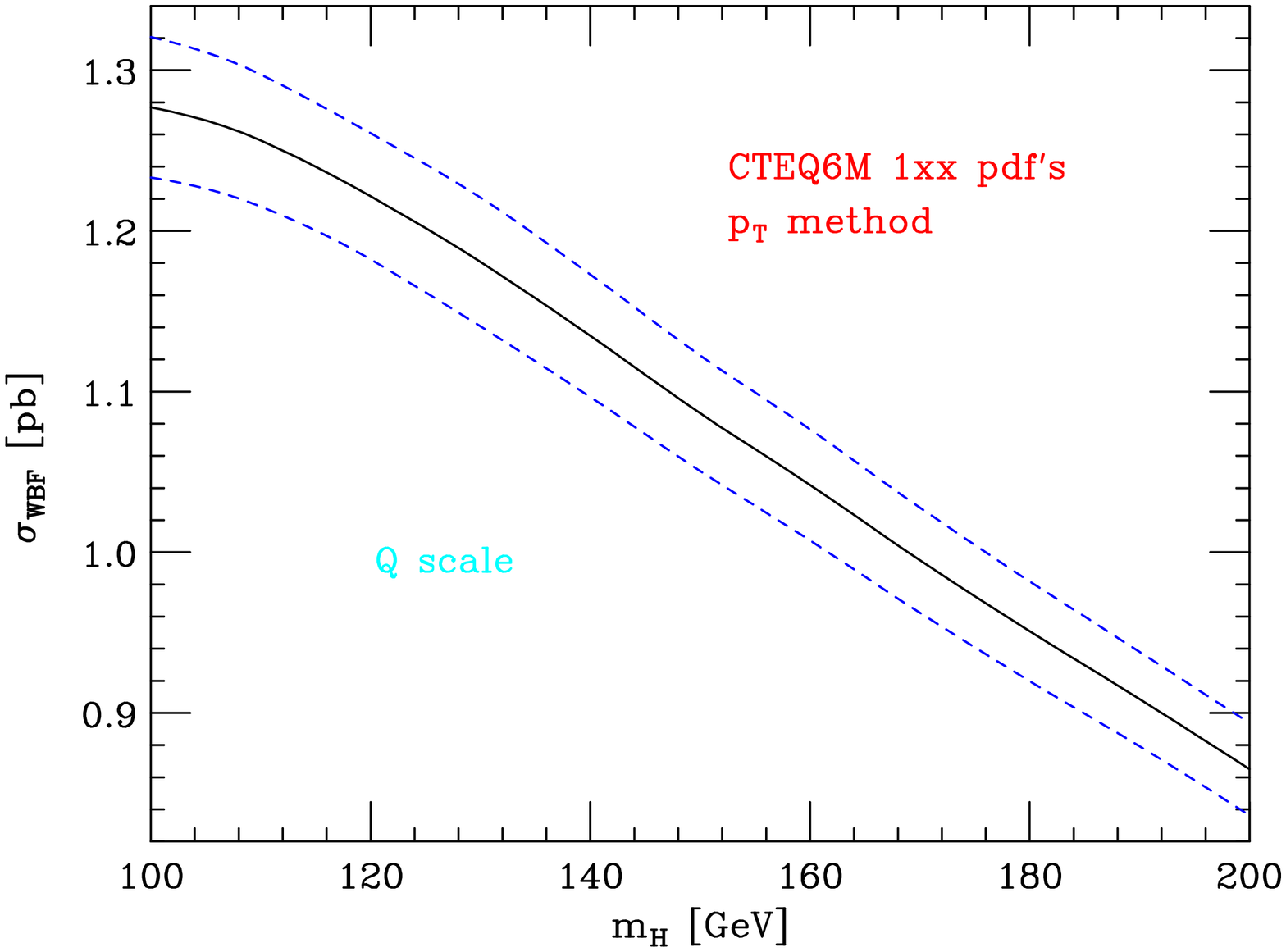,width=0.6\textwidth,angle=90,clip=} \ \  
} 
\ccaption{} 
{\label{fig:pdf_uncert} 
Variation of the total cross section, within cuts, due to errors in the
parton distribution functions, as a function of $m_H$. The central solid line
corresponds to the ``best fit'' CTEQ6M pdf, while the upper and lower curves
define the pdf error band, which is determined from the 40 error eigenvectors
in the CTEQ6M set (CTEQ6M101--CTEQ6M140), adding cross section deviations
in quadrature.}
\end{figure} 

\begin{figure}
\centerline{ 
\epsfig{figure=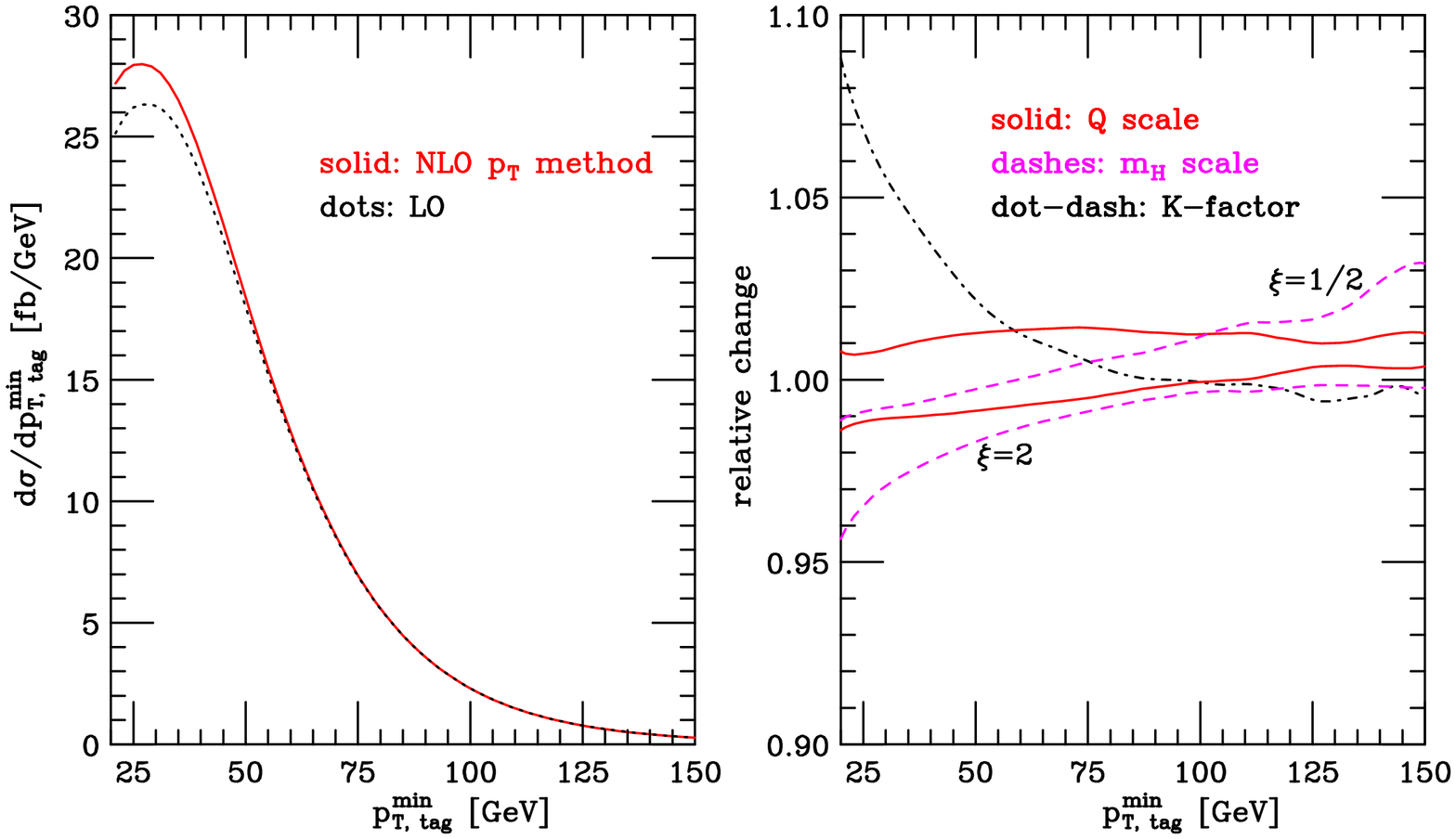,width=0.58\textwidth,angle=90,clip=}
} 
\centerline{ 
\epsfig{figure=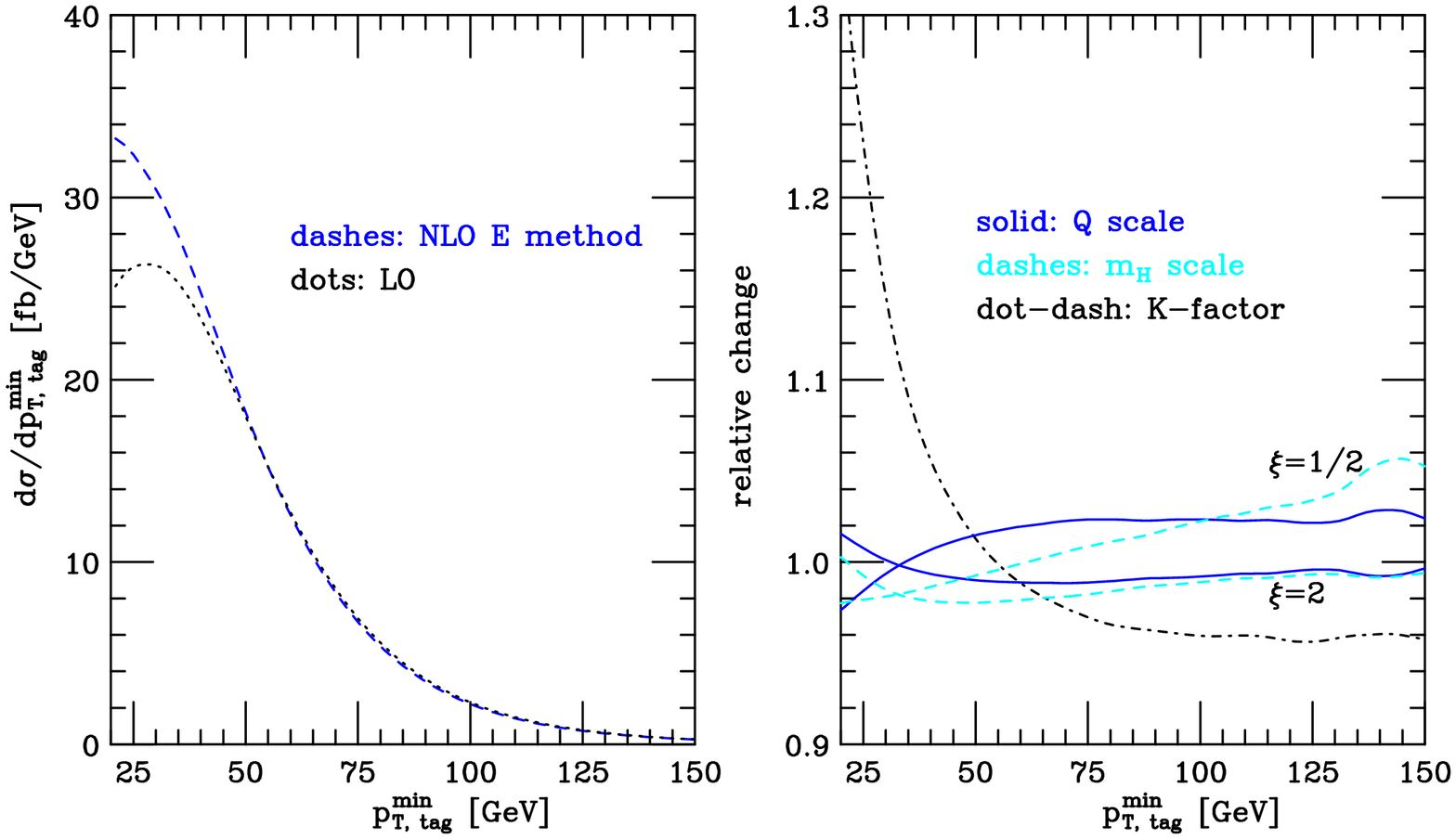,width=0.58\textwidth,angle=90,clip=} 
} 
\vspace*{0.1in}
\ccaption{} 
{ \label{fig:pTjmin} %$d\sigma/dp_{T,\,\tag}^{\rm min}$ (in fb/GeV) 
Transverse momentum distribution of the softer tagging jet for the 
the $p_T$-method (solid red) and the $E$-method (dashed blue) of
defining tagging jets, for $m_H=120$~GeV. 
The right-hand panels give the $K$-factors 
(black dash-dotted line) and the scale variation of the NLO results.
Solid colored curves correspond to $\mu_F=\mu_R=\xi Q_i$ and dashed 
colored curves are for $\mu_F=\mu_R=\xi m_H$ with $\xi=1/2$ and $2$.
}
\end{figure} 

In addition to missing higher order corrections, the theoretical error of the
WBF cross section is dominated by uncertainties in the determination of
the parton distribution functions. We have investigated this 
dependence by calculating the total Higgs boson cross section, within the 
cuts of Eqs.~(\ref{eq:cuts1})--(\ref{eq:cuts4}), for the 40 pdf's in the
CTEQ6Mxxx (xxx = 101--140) set. They correspond to extremal plus/minus
variations in the  
directions of the 20 error eigenvectors of the Hessian of the CTEQ6M fitting
parameters~\cite{cteq6}. Adding the maximum deviations for each error 
eigenvector in quadrature, one obtains the blue dashed lines in 
Fig.~\ref{fig:pdf_uncert}, which define the pdf error band.
We find a uniform $\pm 3.5\%$ pdf uncertainty of the total cross section 
over the entire range of $m_H$ shown.

Scale and pdf uncertainties exhibit little dependence on the Higgs boson mass. 
We therefore limit our investigation to a 
single, representative Higgs boson mass for the remaining discussion, which 
we take as $m_H=120$~GeV.

\begin{figure}
\centerline{ 
\epsfig{figure=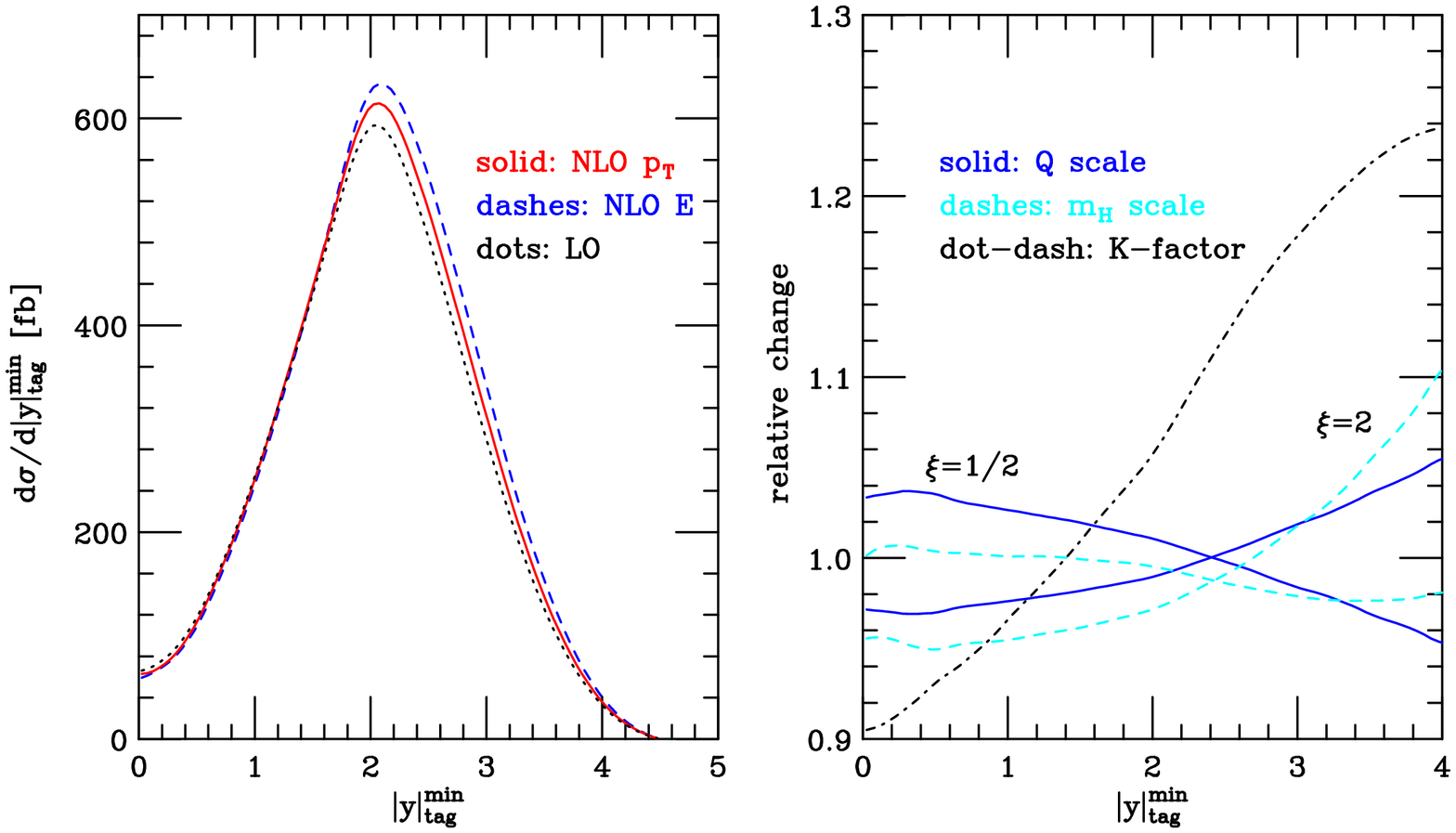,width=0.6\textwidth,angle=90,clip=}
} 
\ccaption{} 
{ \label{fig:y_tag_min} Higgs boson production cross section as a function of
the smaller of the absolute value of the two tagging jet rapidities, $d\sigma/d
|y|_{\tag}^{\rm 
min}$ (in fb, for $m_H=120$~GeV). Results are shown at 
LO (dotted black) and at NLO for the
$p_T$-method (solid red) and the $E$-method (dashed blue) of defining tagging
jets. The right-hand panel gives the $K$-factor (black dash-dotted line) and
the scale variation of the NLO result for the $E$-method.  Colored curves
for the scale dependence are labeled as in Fig.~\ref{fig:pTjmin}.}
\end{figure}

\begin{figure}[thb]
\centerline{ \epsfig{figure=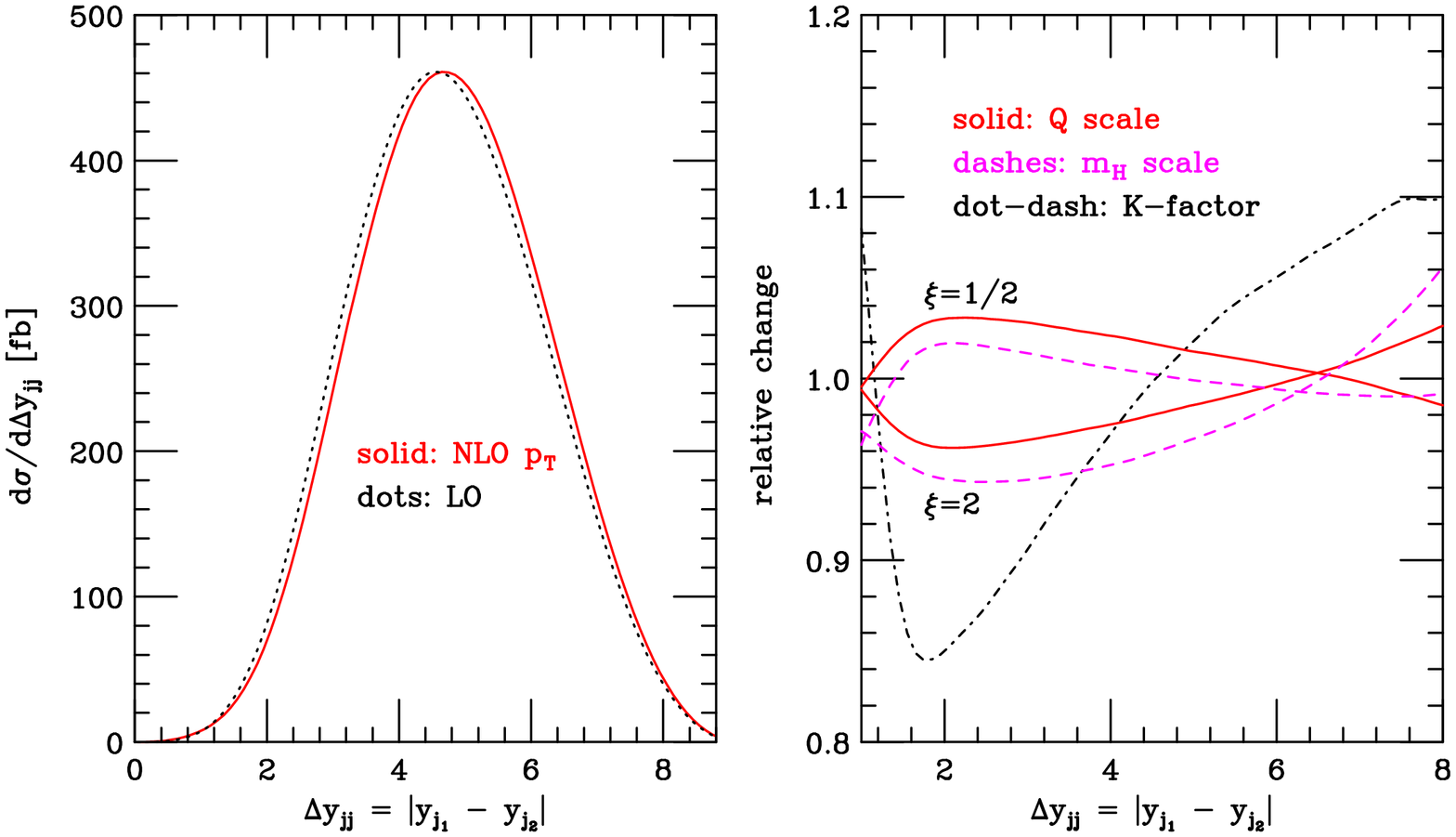,width=0.6\textwidth,angle=90,clip=}
} \ccaption{} { \label{fig:Delyjj} 
Rapidity separation of the two tagging jets for $m_H=120$~GeV.
In the  left-hand panel, $d\sigma/d\Delta y_{jj}$ (in fb) is shown at 
LO (dotted black) and at NLO (solid red), for the $p_T$-method of defining 
tagging jets. The right-hand panel gives the corresponding 
$K$-factor (black dash-dotted line) and the 
scale variation of the NLO results.  Colored curves
for the scale dependence are labeled as in Fig.~\ref{fig:pTjmin}.}
% Solid colored
%curves correspond to $\mu_F=\mu_R=\xi Q$ and dashed colored curves are for
%$\mu_F=\mu_R=\xi m_H$ with $\xi=1/2$ and $2$.  }
\end{figure} 

While the scale dependence of the integrated Higgs boson production cross
section is quite weak, the same need not be true for the shape of
distributions which will be used to discriminate between Higgs boson signal
and various backgrounds.  Having a fully flexible NLO Monte Carlo program at
hand, we can investigate this question. Crucial distributions for the
detection efficiency of the signal are the transverse momentum and the rapidity
of the tagging jets.  In Fig.~\ref{fig:pTjmin} the cross section is shown as
a function of $p_{T,\,\tag}^{\rm min}$, the smaller of the two tagging jet
transverse momenta.  At LO, the tagging jets are uniquely defined, but at NLO
one finds relatively large differences between the $p_T$-method (solid red
curves in the top panels) and the $E$-method (dashed blue curves in bottom
panels). The right-hand-side panels give the corresponding $K$-factors, as
defined in Eq.~(\ref{eq:Kfactor}), (black dash-dotted lines) and the ratio of
NLO differential distributions for different scale choices. Shown are the
ratios 
\beq 
R =\frac{d\sigma^{NLO}(\mu_F=\mu_R=Q_i)}{d\sigma^{NLO}(\mu_F=\mu_R=\mu)} 
\eeq 
for $\mu = 2^{\pm 1}Q_i$ (solid lines) and $\mu = 2^{\pm 1}m_H$ (dashed
lines). While the $K$-factor is modest for the $p_T$-method, it reaches
values around 1.3 in the threshold region for the $E$-method. This strong
rise at NLO is due to hard forward gluon jets being misidentified as tagging
jets in the $E$-method. This problem was recognized previously in parton
shower Monte Carlo simulations and has prompted a preference for the
$p_T$-method~\cite{jakobs}. In spite of the large $K$-factor, however, the
residual scale uncertainty is small, ranging from -4\% to +2\% for the
$p_T$-method and -2\% to +5\% for the $E$-method.

\begin{figure}[thb]
\centerline{ 
\epsfig{figure=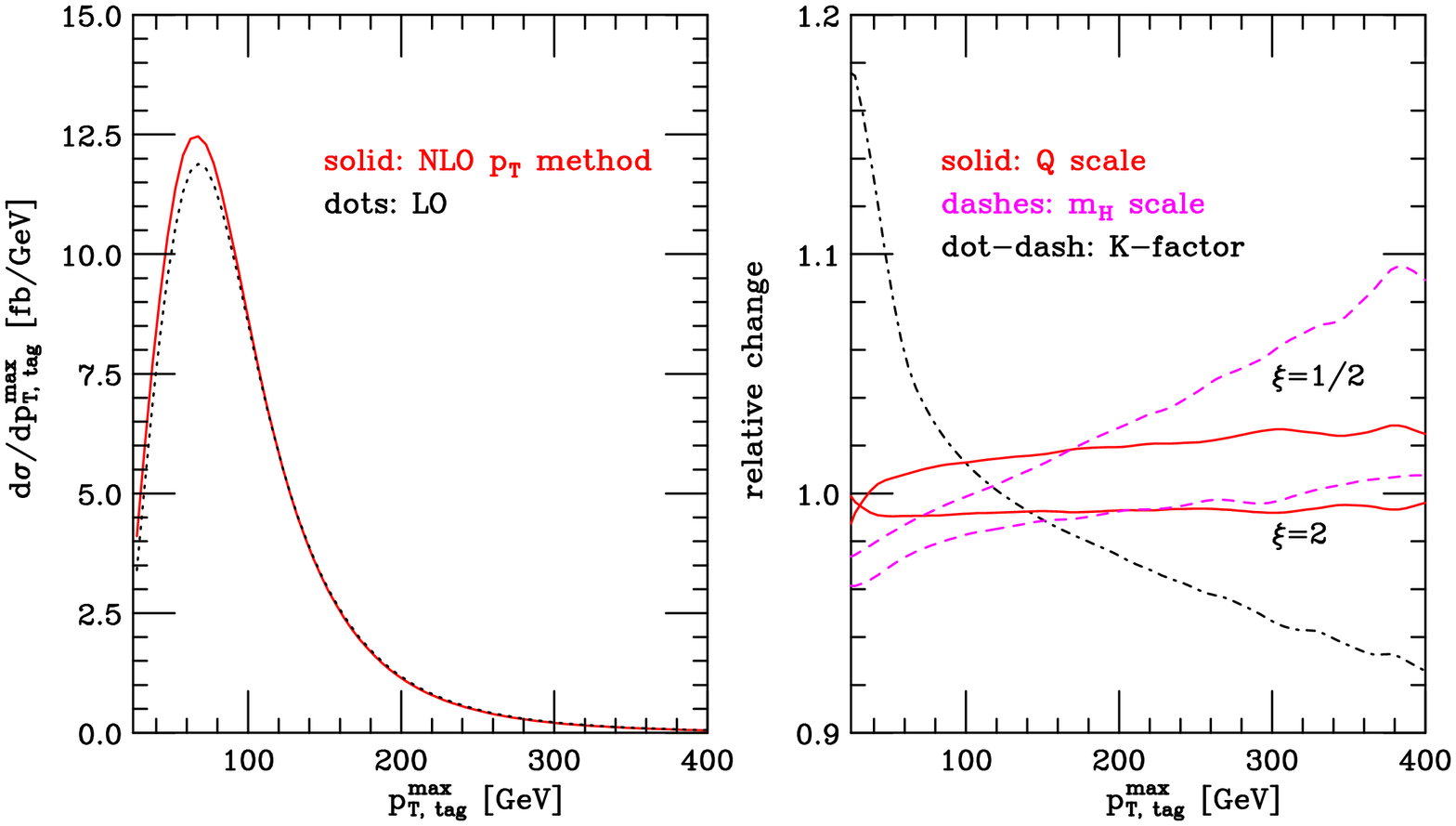,width=0.6\textwidth,angle=90,clip=}
} 
\ccaption{} 
{ \label{fig:pTjmax} 
Transverse momentum distribution of the harder tagging jet, for $m_H=120$~GeV.
In the  left-hand panel, $d\sigma/dp_{T,\,\tag}^{\rm max}$ (in fb/GeV) is shown
at  
LO (dotted black) and at NLO (solid red), for the $p_T$-method of defining 
tagging jets. The right-hand panel gives the corresponding 
$K$-factor (black dash-dotted line) and the 
scale variation of the NLO results.  Colored curves
for the scale dependence are labeled as in Fig.~\ref{fig:pTjmin}.}
%Solid colored curves correspond to $\mu_F=\mu_R=\xi Q$ and dashed 
%colored curves are for $\mu_F=\mu_R=\xi m_H$ with $\xi=1/2$ and $2$.}
\end{figure} 

The more forward selection of tagging jets in the $E$-method is most obvious 
in the rapidity distributions of Figs.~\ref{fig:y_tag_min}
and~\ref{fig:Delyjj}. In Fig.~\ref{fig:y_tag_min} the rapidity of the more  
central of the tagging jets, $|y|_{\rm tag}^{\rm min}$, is shown. At NLO,
the tagging jets are slightly more forward than at tree level, leading to 
a $K$-factor which varies appreciably over phase space. This 
$|y|_{\rm tag}^{\rm min}$-dependence is shown in the right-hand panel for the
$E$-method, together with the residual scale dependence at NLO. Again, scale
variations of less than $\pm 4$\% are found over virtually the entire phase
space. For the $p_T$-method, similar scale variations arise, as
shown in Fig.~\ref{fig:Delyjj} for the rapidity separation between the two
tagging jets, where the cuts of Eqs.~(\ref{eq:cuts1})--(\ref{eq:cuts3}) have 
been imposed.  Figure~\ref{fig:Delyjj} demonstrates that the wide separation 
of the tagging jets, which is important for rejection of QCD backgrounds,
does survive at NLO. In fact, the tagging jet separation even increases 
slightly, making a separation cut like $\Delta y=|y_{j_1}-y_{j_2}|>4$ even 
more effective than at LO.

In all distributions considered so far, no clear preference emerges on whether
to choose the weak-boson virtuality, $Q_i$, or $m_H$ as the hard scale. While
both choices are acceptable, the transverse momentum distributions show 
somewhat smaller scale variations for $\mu=\xi Q_i$ than $\mu=\xi m_H$. The
effect is most pronounced in the high $p_T$ tail of the tagging jet 
distributions. When considering $d\sigma/dp_{T,\rm tag}^{\rm max}$, as
shown in Fig.~\ref{fig:pTjmax}, the scale variation increases to $+10$\%
at large $p_T$ when $\mu=\xi m_H$ is taken, while the same distribution
for  $\mu=\xi Q_i$ stays in a narrow $\pm 2$\% band. This observation
provides another reason for our default scale choice, $\mu = Q_i$.

Unlike the tagging jets considered so far, distributions of the Higgs 
decay products show little change in shape at NLO.

\section{Conclusions}
\label{sec:conclusions}

Weak-boson fusion processes will play an important role at future hadron 
colliders, most notably as a probe for electroweak symmetry breaking. For
the particular case of Higgs boson production, we have presented 
a first analysis of the size and of the remaining uncertainties of NLO
QCD corrections to jet distributions in WBF.  

As for the inclusive WBF cross section, QCD corrections to distributions are
of modest size, of order 10\%, but occasionally they reach larger values. 
These corrections are strongly 
phase-space dependent for jet observables and an overall $K$-factor 
multiplying the LO distributions is not an adequate approximation. Within
the phase-space region relevant for Higgs boson searches, we find differential
$K$-factors as small as 0.9 or as large as 1.3. These corrections need to 
be taken into account for Higgs coupling measurements, and our NLO 
Monte Carlo program, or the recently released analogous program in the MCFM 
package~\cite{MCFM}, provide the necessary tools.

After inclusion of the one-loop QCD corrections, remaining uncertainties due
to as yet uncalculated higher order terms, can be estimated by considering 
scale variations of the NLO cross section. Using the Higgs boson mass, 
$m_H$, and the 
weak-boson virtuality, $Q_i$, as potential hard scales, we find that these
remaining scale dependencies are quite small. Varying renormalization or
factorization scales by a factor of two away from these two central values
results in typical changes of the NLO differential cross sections by
$\pm 2$\% or less. The uncertainty bands for $\mu=\xi m_H$ and $\mu=\xi Q_i$
typically overlap, yielding combined scale uncertainties of less than 
$\pm 3$\% in most cases, occasionally rising up to order 5\% at the edges 
of phase space. Moreover, the variation in different regions typically
cancels in the integrated Higgs cross section, within cuts, leading to
uncertainties due to higher order effects of $\pm 2$\% (see 
Fig.~\ref{fig:sigtot}), even when considering different hard scales.
The remaining theoretical uncertainty on the measurable Higgs cross section,
thus, is well below expected statistical errors, except for the $H\to WW$
search for Higgs masses around 170~GeV, where the high LHC rate allows 
statistical errors as low as 3\%. 
In addition, pdf uncertainties for the total cross section are of order
$\pm3.5\%$ over the range 100~GeV $\le m_H \le$ 200~GeV. This means that 
the SM Higgs boson production cross section via WBF can be predicted, at
present, with a theoretical error of about $\pm 4\%$.

The expected size of the LHC Higgs signal is enhanced slightly by the NLO
QCD corrections. In addition to a $K$-factor slightly above unity this is due
to a small shift of the  tagging jets to higher rapidities, still
well inside the detector coverage but moving the tagging jets slightly farther
apart and hence allowing a better differentiation from QCD backgrounds.

The techniques described in this paper work in a very similar fashion for 
other weak-boson fusion processes. We are planning to extend our work
to include  also $W$ and $Z$ boson 
production via weak-boson fusion, and to make 
these programs generally available. 

\section*{Acknowledgments}

This research was supported in part by the University
of Wisconsin Research Committee with funds granted by the Wisconsin Alumni
Research Foundation and in part by the U.~S.~Department of Energy under
Contract No.~DE-FG02-95ER40896.
C.O. thanks the UK Particle Physics and Astronomy Research Council for
supporting his research.

%\clearpage

%%%%%%%%%%%%%%%%%%%%%%%%%%%%%%%%%%%%%%%%%%%%%%%%%%%%%%%%%%%%%%%%%%%%%%%%%%%%% 


\begin{thebibliography}{99}
\bibitem{CMS}
G.~L.~Bayatian {\it et al.}, CMS Technical Proposal,
report CERN/LHCC/94-38x (1994);
%D. Denegri, %Prospects for Higgs (SM and MSSM) searches at LHC,
%talk in the Circle Line Tour Series, Fermilab, October 1999,
%(http://www-theory.fnal.gov/CircleLine/DanielBG.html);
R.~Kinnunen and D.~Denegri, %Expected SM/SUSY Higgs observability in CMS,
CMS NOTE 1997/057;
R. Kinnunen and A. Nikitenko,
%Study of $H_{SUSY}\to \tau\tau\to l^{\pm}+h^{\mp} + E_t^{miss}$ in CMS,
CMS TN/97-106;
R.~Kinnunen and D.~Denegri,
%The $H_{SUSY}\to \tau\tau\to h^{\pm}+h^{\mp}+X$
%channel, its advantages and potential instrumental drawbacks,
arXiv:hep-ph/9907291;
%%CITATION = HEP-PH 9907291;%%
V.~Drollinger, T.~M\"uller and D.~Denegri,
arXiv:hep-ph/0111312.
%%CITATION = HEP-PH 0111312;%%

\bibitem{ATLAS}
ATLAS Collaboration, ATLAS TDR,
%ATLAS Detector and Physics Performance Technical Design Report,
report CERN/LHCC/99-15 (1999);
E.~Richter-Was and M.~Sapinski,
Acta Phys.\ Polon.\ B {\bf 30}, 1001 (1999);
%%CITATION = APPOA,B30,1001;%%
B.~P.~Kersevan and E.~Richter-Was,
Eur.\ Phys.\ J.\ C {\bf 25}, 379 (2002)
[arXiv:hep-ph/0203148].
%%CITATION = HEP-PH 0203148;%%

\bibitem{Zeppenfeld:2000td}
D.~Zeppenfeld, R.~Kinnunen, A.~Nikitenko and E.~Richter-Was,
%``Measuring Higgs boson couplings at the LHC,''
Phys.\ Rev.\ {\bf D62}, 013009 (2000)
[arXiv:hep-ph/0002036];
%%CITATION = HEP-PH 0002036;%%
D.~Zeppenfeld,
%``Higgs couplings at the LHC,''
in {\it Proc. of the APS/DPF/DPB Summer Study on the Future of 
Particle Physics (Snowmass 2001) } ed. N.~Graf,
eConf {\bf C010630}, P123 (2001)
[arXiv:hep-ph/0203123];
%%CITATION = HEP-PH 0203123;%%
A.~Belyaev and L.~Reina,
%``p p $\to$ t anti-t H, H $\to$ tau+ tau-: Toward a model independent  
%determination of the Higgs boson couplings at the LHC,''
JHEP {\bf 0208}, 041 (2002)
[arXiv:hep-ph/0205270].
%%CITATION = HEP-PH 0205270;%%

\bibitem{wbfhtautau}
D.~Rainwater, D.~Zeppenfeld and K.~Hagiwara,
%``Searching for H $\to$ tau tau in weak boson fusion at the LHC,''
Phys.\ Rev.\ D {\bf 59}, 014037 (1999) [arXiv:hep-ph/9808468];
%%CITATION = HEP-PH 9808468;
T.~Plehn, D.~Rainwater and D.~Zeppenfeld,
%``A method for identifying H --> tau tau --> e+- mu-+ missing p(T)
%at the CERN LHC,''
Phys.\ Rev.\  {\bf D61}, 093005 (2000) [arXiv:hep-ph/9911385];
%[hep-ph/9911385].
S.~Asai et al., ATL-PHYS-2003-005.

\bibitem{wbfhtoww}
D.~Rainwater and D.~Zeppenfeld,
%``Observing $H \to W^{(*)}W^{(*)} \to e^\pm \mu^\mp /\!\!\!{p}_T$ in weak boson fusion with dual forward jet tagging at the CERN LHC,''
Phys.\ Rev.\ D {\bf 60}, 113004 (1999)
[Erratum-ibid.\ D {\bf 61}, 099901 (2000)]
[arXiv:hep-ph/9906218];
%%CITATION = HEP-PH 9906218;%%
N.~Kauer, T.~Plehn, D.~Rainwater and D.~Zeppenfeld,
%``H $\to$ W W as the discovery mode for a light Higgs boson,''
Phys.\ Lett.\ B {\bf 503}, 113 (2001)
[arXiv:hep-ph/0012351];
%%CITATION = HEP-PH 0012351;%%
C.~M.~Buttar, R.~S.~Harper and K.~Jakobs,
ATL-PHYS-2002-033;
K.~Cranmer et al.,
ATL-PHYS-2003-002 and ATL-PHYS-2003-007;
S.~Asai et al.,
ATL-PHYS-2003-005.

\bibitem{wbfhtophoton}
D.~Rainwater and D.~Zeppenfeld,
%``Searching for H $\to$ gamma gamma in weak boson fusion at the LHC,''
JHEP {\bf 9712}, 005 (1997)
[arXiv:hep-ph/9712271].
%%CITATION = HEP-PH 9712271;%%


\bibitem{Eboli:2000ze}
O.~J.~Eboli and D.~Zeppenfeld,
%``Observing an invisible Higgs boson,''
Phys.\ Lett.\ B {\bf 495}, 147 (2000)
[arXiv:hep-ph/0009158];
%%CITATION = HEP-PH 0009158;%%
B.~Di Girolamo, A.~Nikitenko, L.~Neukermans, K.~Mazumdar and D.~Zeppenfeld,
%``Experimental observation of an invisible Higgs boson at LHC,''
in arXiv:hep-ph/0203056.

\bibitem{WBF_NLO}
T.~Han and S.~Willenbrock, Phys.\ Lett.\ {\bf B273}, 167 (1991).
%%CITATION = PHLTA,B273,167;%%

\bibitem{HggNLO}
A.~Djouadi, N.~Spira and P.~Zerwas, Phys. Lett. {\bf B264}, 440 (1991);
M.~Spira, A.~Djouadi, D.~Graudenz and P.M.~Zerwas,
%``Higgs boson production at the LHC,''
Nucl.\ Phys.\ {\bf B453}, 17 (1995);
%%CITATION = NUPHA,B453,17;%%
S. Dawson, Nucl.~Phys. {\bf B359}, 283 (1991);
%%CITATION = NUPHA,B359,283;%%
S.~Catani, D.~de Florian, M.~Grazzini and P.~Nason, in arXiv:hep-ph/0204316.

\bibitem{H2loop}
S.~Catani, D.~de Florian and M.~Grazzini, JHEP {\bf 0105}, 025  (2001)
[arXiv:hep-ph/0102227];
%%CITATION = HEP-PH 0102227;%%
R.~Harlander and W.~Kilgore, Phys.\ Rev.\ {\bf D64}, 013015  (2001)
[arXiv:hep-ph/0102241];
%%CITATION = HEP-PH 0102241;%%
%
R.~Harlander and W.~Kilgore, Phys.\ Rev.\ Lett.\ {\bf 88}, 201801 (2002)
[arXiv:hep-ph/0201206];
%%CITATION = HEP-PH 0201206;%% 
%
C.~Anastasiou and K.~Melnikov, Nucl.\ Phys.\ {\bf B646}, 220 (2002)
[arXiv:[hep-ph/0207004];
%%CITATION = HEP-PH 0207004;%% 
%
V.~Ravindran, J.~Smith and W.L.~van Neerven,  arXiv:hep-ph/0302135.
%%CITATION = HEP-PH 0302135;%% 

\bibitem{CS}
S.~Catani and M.~H.~Seymour,
%``A general algorithm for calculating jet cross sections in NLO QCD,''
Nucl.\ Phys.\ B {\bf 485}, 291 (1997)
[Erratum-ibid.\ B {\bf 510}, 503 (1997)]
[arXiv:hep-ph/9605323].
%%CITATION = HEP-PH 9605323;%%

\bibitem{HZ}
K.~Hagiwara and D.~Zeppenfeld,
%``Helicity Amplitudes For Heavy Lepton Production In E+ E- Annihilation,''
Nucl.\ Phys.\ B {\bf 274}, 1 (1986);
%%CITATION = NUPHA,B274,1;%%
K.~Hagiwara and D.~Zeppenfeld,
%``Amplitudes For Multiparton Processes Involving A Current ...,''
Nucl.\ Phys.\ B {\bf 313}, 560 (1989).
%%CITATION = NUPHA,B313,560;%%

\bibitem{vegas}
G.~P. Lepage, J. Comput. Phys. {\bf 27},  192  (1978).

\bibitem{Spira:1997dg}
M.~Spira,
%``QCD effects in Higgs physics,''
Fortsch.\ Phys.\  {\bf 46}, 203 (1998)
[arXiv:hep-ph/9705337].
%%CITATION = HEP-PH 9705337;%%

\bibitem{cteq6}
J.~Pumplin, D.~R.~Stump, J.~Huston, H.~L.~Lai, P.~Nadolsky and W.~K.~Tung,
%``New generation of parton distributions with uncertainties ...,''
JHEP {\bf 0207}, 012 (2002)
[arXiv:hep-ph/0201195].
%%CITATION = HEP-PH 0201195;%%

\bibitem{kToriginal}
S.~Catani, Yu.~L.~Dokshitzer and B.~R.~Webber,
                Phys.\ Lett.\ B  {\bf 285} 291 (1992);
S.~Catani, Yu.~L. Dokshitzer, M.~H.~Seymour and B.~R.~Webber,
                Nucl.\ Phys.\ B {\bf 406} 187 (1993);
S.~D.~Ellis and D.~E.~Soper, Phys.\ Rev.\,
{\bf D48} 3160 (1993).

\bibitem{kTrunII}
G.~C.~Blazey {\it et al.},
%``Run II jet physics,''
arXiv:hep-ex/0005012.
%%CITATION = HEP-EX 0005012;%%

\bibitem{jakobs}
V.~Cavasinni, D.~Costanzo, E.~Mazzoni and I.~Vivarelli,
ATL-PHYS-2002-010.
%C.~M.~Buttar, R.~S.~Harper and K.~Jakobs,
%ATL-PHYS-2002-033.

\bibitem{MCFM}
J.~Campbell and K.~Ellis, MCFM - Monte Carlo for FeMtobarn processes,
http://mcfm.fnal.gov/.

\end{thebibliography}
\end{document}